\documentclass[pre,twocolumn,groupedaddress,showkeys]{revtex4-1}
\usepackage{graphicx}

\begin{document}

\title{Noise-induced Synchronization of Crystal Oscillators}

\author{Kazuyoshi Ishimura}
\email[]{kazu-i@fc.ritsumei.ac.jp}
\author{Isao T. Tokuda}
\email[]{isao@fc.ritsumei.ac.jp}
\affiliation{Department of Mechanical Engineering, 
Ritsumeikan University, 1-1-1 Noji-Higashi, Kusatsu, 
Shiga 525-8577, Japan}

\date{\today}

\begin{abstract}
Experimental study on noise-induced synchronization of crystal 
oscillators is presented. 
Two types of circuits were constructed: 
one consists of two \textit{Pierce} oscillators that were isolated 
from each other and received a common noise input,
while the other is based on a single \textit{Pierce} oscillator that 
received a same sequence of noise signal repeatedly.
Due to frequency detuning between the two \textit{Pierce} oscillators, 
the first circuit showed no clear sign of noise-induced synchronization.
The second circuit, on the other hand, generated coherent waveforms 
between different trials of the same noise injection. The waveform 
coherence was, however, broken immediately after the noise injection 
was terminated. 
Stronger modulation such as the voltage resetting was finally shown
to be effective to induce phase shifts, leading to 
phase--synchronization of the \textit{Pierce} oscillator. 
Our study presents a guideline for synchronizing clocks of multiple 
CPU systems, distributed sensor networks, and other engineering devices.
\end{abstract}

\pacs{XX}

\maketitle
\section{Introduction}
Synchronization is a ubiquitous phenomenon of 
coupled nonlinear oscillators found in a diverse fields of science 
and engineering.
During the last several decades, remarkable progress has been made 
in both theory and experiment on synchronization of limit cycle 
and chaotic oscillators
\cite{winfree2001,kuramoto1984,pikovsky2003synchronization}.
Although the most common situation presumes that the oscillators 
are directly coupled with each other, indirect interaction through 
commonly injected noise sources has been also known to synchronize 
uncoupled oscillators.
It has been proven theoretically that a wide class of uncoupled 
limit-cycle oscillators can be in-phase synchronized by common weak 
white noise \cite{teramae2004robustness}.
The theoretical framework has been extended to random impulses 
\cite{arai2008commonrandomimpulse,nagai2009commontelegraphnoise}, 
general, colored and \textit{non-Gaussian}, noise 
\cite{goldobin2010generalnoise}, 
and also in the presence of uncommon noise 
\cite{nakao2007noise}.
Chaotic oscillators can be also synchronized by noise inputs
\cite{lai1998synchronization,rim2000chaotic,zhou2002noiseinduced,yoshimura2007synchronization}.

Experimental systems that show the noise-induced synchronization 
include neuronal systems 
\cite{mainen1995reliability,de2007correlation,galan2006correlation}, 
circuit systems 
\cite{nagai2009commontelegraphnoise,arai2008phase,yoshida2006noise} 
and laser systems \cite{sunada2014optical,tomiyama2018effect}.
Despite intensive studies on noise-induced synchronization in 
a scientific framework, its application to engineering problems 
remains open.
Natural environmental sounds have been utilized as a noise
source to synchronize simulated network of distributed 
sensors \cite{yasuda2013natural}.
Numerical study showed that noise can synchronize spin torque 
oscillators to overcome the problem of low output power 
in their array \cite{nakada2012noise}.

In the present study, we apply the method of noise-induced 
synchronization to crystal oscillators
\cite{vittoz1988high,matthys1983crystal}.
The crystal oscillator is an electronic circuit composed of 
a piezoelectric resonator that determines the oscillation 
frequency. 
Because of its highly precise and stable oscillations, 
it has been widely used to provide clock signals to 
a variety of digital circuits and 
to stabilize frequencies of radio transmitters and receivers. 
The clock rate of a central processing unit (CPU) is also
determined by the frequency of the crystal oscillator.
Under these circumstances, synchronized operation of multiple 
clocks should be of significant importance in a future technology. 
For instance, to further accelerate the clock speed in CPUs, 
it is getting extremely difficult to integrate all the circuit 
elements into a single chip, because of the increased number 
of transistors, which should be located within 
a limited-size chip. 
Diving the CPU into multiple chips is inevitable, thereby
synchronizing the clocks of the divided chips should be 
an essential requirement.
As a source of the noise to synchronize the oscillators,
utilization of the internal noise, which exists inherently 
within circuit itself, should be advantageous in terms of
an economical recycling of the oscillation energy. 
A variety of further applications, \textit{e.g.}, synchronizing 
the clocks of distributed sensors and CPUs, should be found.
Towards establishment of a basis for such technologies, 
we present here an experimental study on noise-induced 
synchronization of the crystal oscillators.
Two types of circuits were constructed:
(1) Two \textit{Pierce} oscillators that were isolated from each other 
and received a common noise input; 
(2) A single \textit{Pierce} oscillator that repeatedly received
a same noise input.
As the external input signals, white \textit{Gaussian} noise, 
\textit{Poisson} spike trains, and reseting signals were applied 
to the circuits.
Our study may provide a guideline for synchronizing clocks of multiple 
CPU systems, distributed sensor networks, and other engineering devices.

The present paper is organized as follows. 
Section II briefly introduces the theoretical framework for noise-induced
synchronization of limit cycle oscillators. 
Section III implements two isolated \textit{Pierce} oscillator circuits, 
to which common noise inputs were applied. 
Section IV applies a same noise input repeatedly to a single \textit{Pierce} 
oscillator circuit to examine whether coherence of the output waveforms 
is increased by the same noise injection. 
Section V examines the effect of voltage resetting on the phase dynamics
of the \textit{Pierce} circuit.
The final Section is devoted to conclusions and discussions of this study.

\section{Theory of Noise-induced Synchronization}
We consider two identical non-interacting limit cycle oscillators 
driven by a common noise 
\begin{eqnarray}
\frac{dX_i}{dt} = F(X_i) + {\epsilon} {\xi}(t) e,
\label{eqn:drivenlimitcycle}
\end{eqnarray}
where $i=1,2$ are indices for the two oscillators and ${\xi}(t)$ 
represents \textit{Gaussian} white noise with
${\langle}{\xi}(t){\rangle} = 0$ and
${\langle}{\xi}(t) {\xi}(s){\rangle} = {\delta}(t-s)$
(${\langle}{\cdot}{\rangle}$ represents statistical average).
It is assumed that the noise is applied only in the direction $e$ 
in the state space. 
By applying the phase reduction \cite{kuramoto1984}, 
the oscillator dynamics is reduced to the following phase equation
\begin{eqnarray}
\frac{d}{dt} {\theta}_{i}(t) = {\omega} + {\epsilon} Z({\theta}_{i}(t)) {\xi}(t),
\label{eqn:phasemodel}
\end{eqnarray}
where $Z({\theta})$ represents phase sensitivity function of 
the oscillator with respect to perturbation applied in the 
direction $e$. Denoting the phase difference by
${\phi}(t) = {\theta}_{1}(t) - {\theta}_{2}(t)$, 
subtraction of the two phase equations yields 
\begin{eqnarray}
\frac{d}{dt} {\phi}(t) &=& {\epsilon} {\{} Z({\theta}_{1}(t)+{\phi}(t))
- Z({\theta}(t)) {\}} {\xi}(t)
\nonumber\\
&=& {\epsilon} Z'({\theta}(t)){\phi}(t){\xi}(t), 
\label{eqn:phasedifference}
\end{eqnarray}
where the phase sensitivity function was expanded as 
$Z({\theta}+{\phi}) = Z({\theta}) + Z'({\theta}) {\phi} + O({\phi}^{2})$
for sufficiently small phase difference ${\vert} {\phi} {\vert}$.
The logarithm of the absolute phase difference 
${\ln} {\vert} {\phi}(t) {\vert}$ therefore obeys 
\begin{eqnarray}
\frac{d}{dt}  {\ln} {\vert} {\phi}(t) {\vert}
= {\epsilon} Z'({\theta}(t)) {\xi}(t).
\label{eqn:logphasedifference}
\end{eqnarray}
Thus, growth rate of ${\ln} {\vert} {\phi}(t) {\vert}$ is 
determined by the mean Lyapunov exponent
${\Lambda} = {\langle} \frac{d {\ln} {\vert} {\phi}(t) {\vert}}{dt} {\rangle}$,
which is statistically averaged over noise.
For the \textit{Gaussian} white noise, the Lyapunov exponent 
is calculated as
${\Lambda} = - \frac{{\epsilon}^{2}}{4{\pi}} 
{\int}_{0}^{2{\pi}} {\{} Z'({\theta}) {\}}^{2} d{\theta} 
{\leq} 0$, 
implying that 
${\vert} {\phi}(t) {\vert}$ shrinks on average and the oscillators 
get eventually synchronized with each other. 
This holds for arbitrary limit cycle oscillators regardless of 
the detailed dynamics as long as $Z$ is differentiable
\cite{teramae2004robustness}.
The theoretical framework has been extended to random impulses 
\cite{arai2008commonrandomimpulse,nagai2009commontelegraphnoise}, 
general, colored and \textit{non-Gaussian}, noise \cite{goldobin2010generalnoise}, 
in the presence of uncommon noise \cite{nakao2007noise},
and also to chaotic oscillators \cite{yoshimura2007synchronization}.

The crystal circuit system generates limit cycle oscillations, which 
fall within the above theoretical framework. 
In the following Sections, we experimentally apply common noise inputs
to the crystal oscillator system. 

\section{Two Uncoupled Pierce Circuits}
\begin{figure}[t!]
\centering
(a)\includegraphics[scale=0.50]{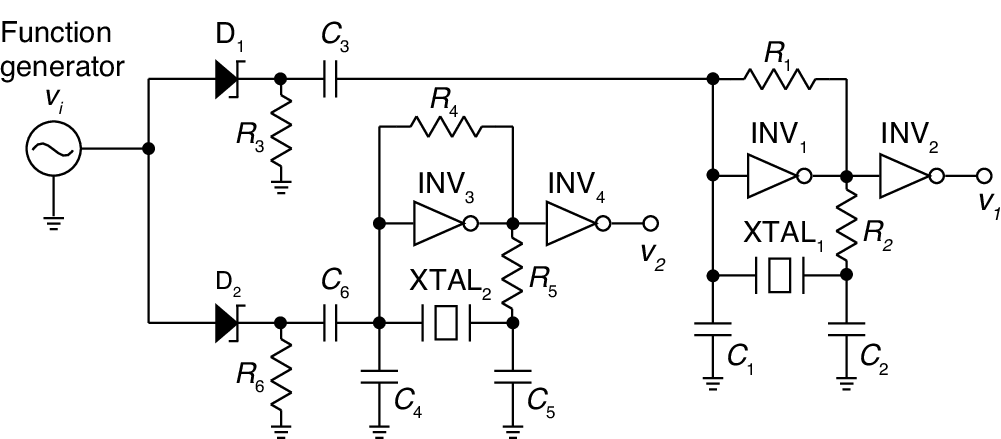}
(b)\includegraphics[scale=0.28]{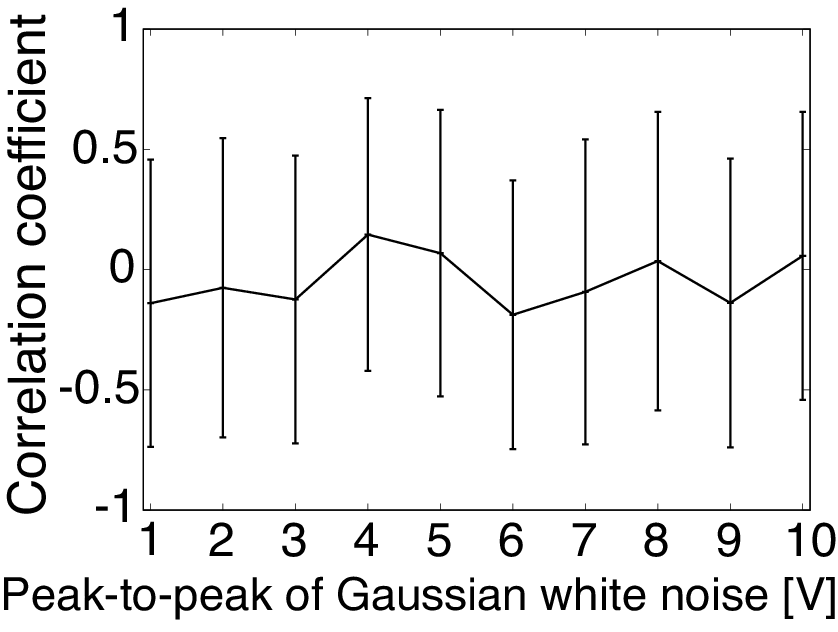}
(c)\includegraphics[scale=0.28]{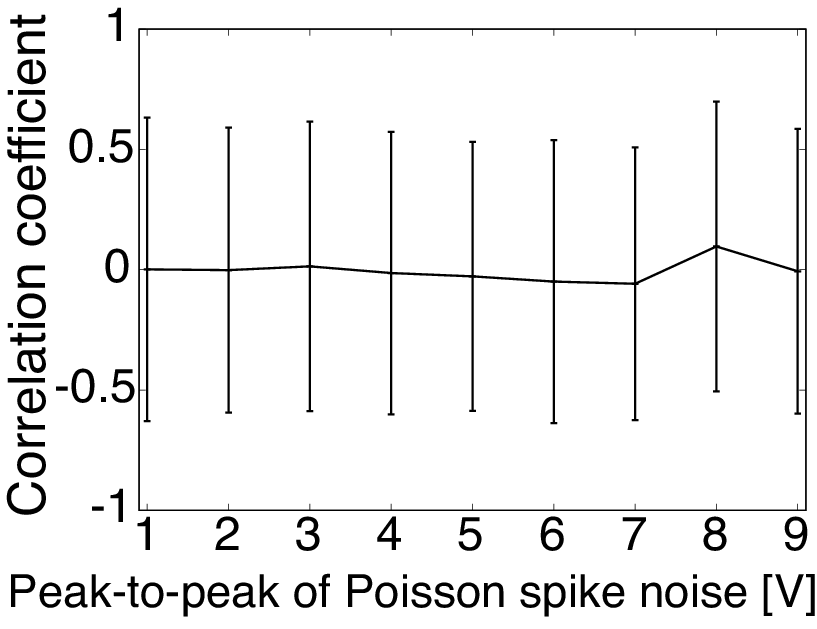}
(d)\includegraphics[scale=0.28]{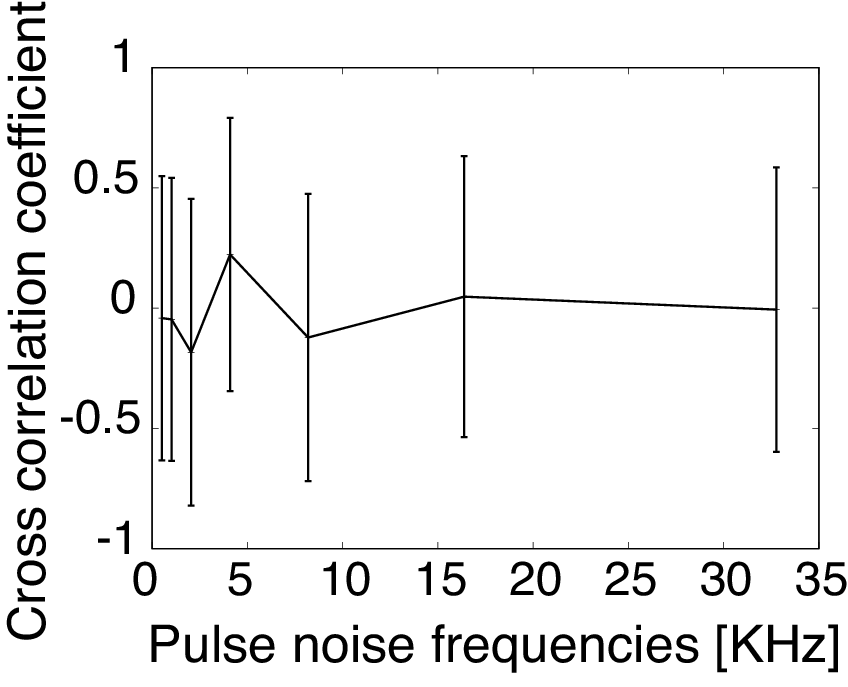}
(e)\includegraphics[scale=0.28]{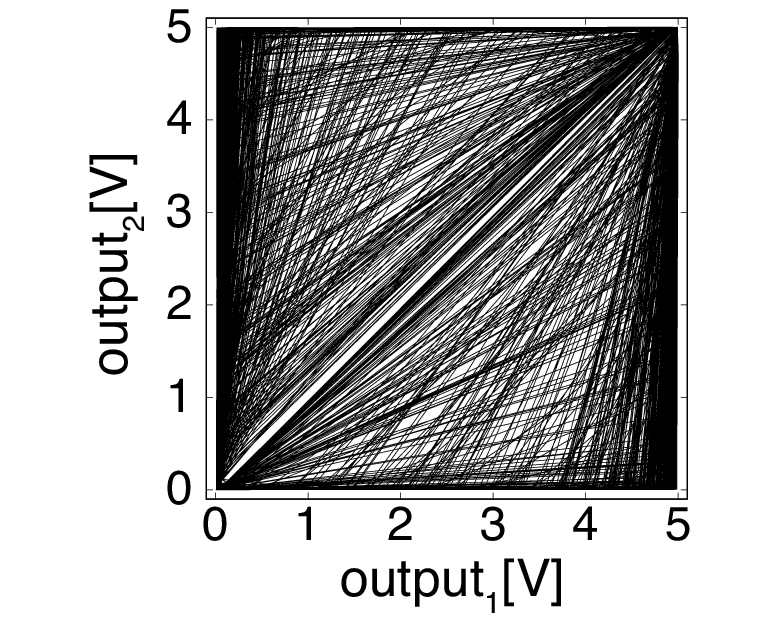}
\caption{
(a) Schematic diagram of two uncoupled \textit{Pierce} oscillator 
circuits. Each circuit consists of inverting operational amplifier 
circuit (${\rm INV_1}$ or ${\rm INV_2}$), 
feedback resistor ($R_1$ or $R_4$), 
crystal (${\rm XTAL_1}$ or ${\rm XTAL_2}$), and 
capacitors ($C_1$, $C_2$ or $C_4$, $C_5$).
The function generator (Keysight 33500B) injects 
external signals 
(\textit{Gaussian} white noise or \textit{Poisson} spike trains)
to each oscillator via diode (${\rm D_1}$ or ${\rm D_2}$), 
coupling capacitor ($C_1$ or $C_4$), and 
ground resistance ($R_2$ or ${R_4}$).
(b) Dependence of the correlation coefficient between $V_{1}$ 
and $V_{2}$ on the peak-to-peak voltage of \textit{Gaussian} 
white noise input. 
Averaged value over 10 different measurements is plotted with 
the standard deviation as the error-bar.
(c) Dependence of the correlation coefficient on the amplitude 
of \textit{Poisson} spike trains. 
The mean spike frequency was approximately 32.768 kHz, 
which is close to the natural frequency of crystals. 
(d) Dependence of the correlation coefficient on frequencies of 
the \textit{Poisson} spike inputs (amplitude: 9 V).
(e) {\it Lissajous} plot of $V_{1}$ \textit{vs.} $V_{2}$ 
in the case of \textit{Poisson} spike inputs 
(frequency: 32.768 kHz, amplitude: 9 V).
\textit{Poisson} spike noise.
}
\label{two_pierce}
\end{figure}

As the piezoelectric crystal oscillator circuits, two \textit{Pierce} 
oscillators \cite{pierce1923piezoelectric}, each of which was composed 
of a single digital inverter (Toshiba 74HCU04AP), one resistor, 
two capacitors, and one quartz crystal ($32.768$ kHz, SII VT-200-F), 
were built as shown in Fig.~\ref{two_pierce} (a).
The inverters (${\rm INV_1}$ and ${\rm INV_3}$) amplified oscillations 
of the crystals (${\rm XTAL_1}$ and ${\rm XTAL_2}$) with feedback 
resistors ($R_1$ and $R_4$), while the capacitors ($C_1$, $C_2$, $C_4$, 
and $C_5$) adjusted their oscillation frequencies.
The output signals, denoted as $V_{1}$ and $V_{2}$, respectively, 
were generated through the buffers (${\rm INV_2}$ and ${\rm INV_4}$).
To isolate the two \textit{Pierce} oscillators from each other,
Schottky diodes ($D_1$ and $D_2$) that rectified their inputs were 
inserted. 
The coupling capacitors ($C_3$ and $C_6$) passed the alternating input 
signals, while the resistors ($R_3$ and $R_6$) discharged them. 
Physical parameters of the electric components used in the present 
experiment are summarized in Table~\ref{piercec}. 
To inject a common noise input, a function generator (Keysight 33500B), 
which generates various noise signals including white \textit{Gaussian} 
noise and \textit{Poisson} spike trains, was utilized. 
The output voltages ($V_{1}$, $V_{2}$) were simultaneously observed 
by an oscilloscope (Keysight InfiniiVision DSOX2014A) and 
recorded into a data logger (Keyence NR-600 and NR-HA08) 
with a sampling frequency of $f_{s}=500$ kHz and 
with a data point of $N=2{\times}10^{6}$. 

To quantify the level of synchronization between the two crystal 
oscillators, correlation coefficient between the two outputs, 
$V_{1}$ and $V_{2}$, was computed.
To study dependence of the synchronization on the noise strength, 
peak-to-peak voltage of the input noise was increased from 0 to 10 V.
Fig.~\ref{two_pierce} (b) shows the results of applying \textit{Gaussian} 
white noise.
For each setting of the noise strength, $10$ measurements were repeatedly
made with different realizations of the noise signals and their average 
was drawn with the standard deviation indicated by the error-bar. 
The correlation coefficient remains close to zero even if the noise strength 
is increased to the maximum level of 10 V.
Fig.~\ref{two_pierce} (c), on the other hand, shows the results of 
applying \textit{Poisson} spike trains with an average spike frequency 
of 32.768 kHz, which is close to the natural frequency of the crystal 
oscillator. Again, the correlation coefficient stays close to the
zero line up to the maximum spike amplitude of 9 V. 
The results were the same for different setting of the spike frequency,
which was varied as ??, with the peak-to-peak voltage set to 9 V 
(Fig.~\ref{two_pierce}~(d)). 

To visualize the inter--relationship between the two \textit{Pierce} 
circuits, {\it Lissajous} plot of $V_{1}$ \textit{vs.} $V_{2}$ was 
drawn for the \textit{Poisson} spike input (amplitude: 9 V, average 
frequency: 32.768 kHz) in Fig.~\ref{two_pierce} (e). 
The phase plots, which filled the entire square region, showed 
no sign of coherence. 
This indicates that the noise-induced synchronization was not 
observed for the present two uncoupled crystal oscillator circuits.
As shown in \cite{yoshimura2007synchronization}, noise-induced 
synchronization is, in a strict sense, not achievable in the 
presence of frequency detuning between the two limit cycle oscillators.
To observe synchronization, frequency mismatch, which inevitably 
exists between the two crystal oscillators, should be largely 
reduced.

\begin{table}[t]
\centering
\caption{Parameters of Two Pierce Circuits.}
\label{piercec}
  \begin{tabular}{@{\hspace{2em}}c@{\hspace{1.5em}}|l@{\hspace{2em}}}
    \multicolumn{2}{c}{Circuit elements of \textit{Pierce} oscillators}\\ \hline
${\rm XTAL_1}$ and ${\rm XTAL_2}$ & $32.768$ KHz\\ 
${\rm INV_1}$, ${\rm INV_2}$, ${\rm INV_3}$ and ${\rm INV_4}$ & 74HCU04AP\\
$R_1$ and $R_3$ & $1$ M$\Omega$\\
$C_2$, $C_3$, $C_5$ and $C_6$ & $22$ pF\\ \hline
    \multicolumn{2}{c}{Unidirectional input paths}\\ \hline
${\rm D_1}$ and ${\rm D_2}$ & BAT43\\
$R_3$ and $R_4$ & $10$ k$\Omega$\\
$C_1$ and $C_4$ & $20$ pF\\
    \hline
  \end{tabular}
\end{table}

\section{Single Pierce Circuit with Repeated Noise Injection}
In the previous Section, we have seen that the noise-induced 
synchronization was not observed for the two uncoupled crystal 
circuit systems. The frequency mismatch, which inherently exists 
between the two circuits, is considered as one of the primary 
causes that prevented the synchrony. 
In order to exclude such primary cause, here, we utilize 
a single \textit{Pierce} oscillator circuit.
By repeatedly injecting a same noise sequence to the circuit, 
we may measure the coherence of the circuit outputs between 
different trials. 
By considering that each trial output is generated from different
circuit having exactly the same oscillator properties, noise-induced
synchronization can be examined between identical oscillators 
that have no frequency mismatch,
as demonstrated in \cite{mainen1995reliability}.

\begin{figure}[t!]
\centering
(a)\includegraphics[scale=0.5]{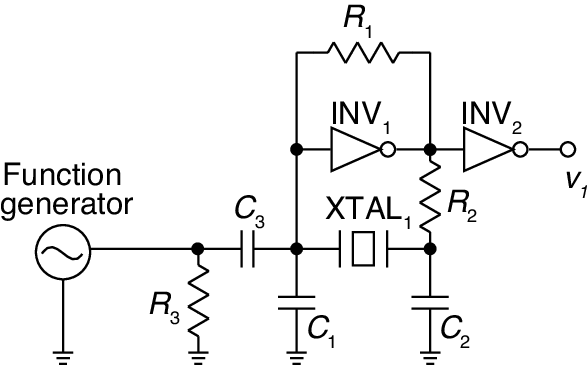}
(b)\includegraphics[scale=0.28]{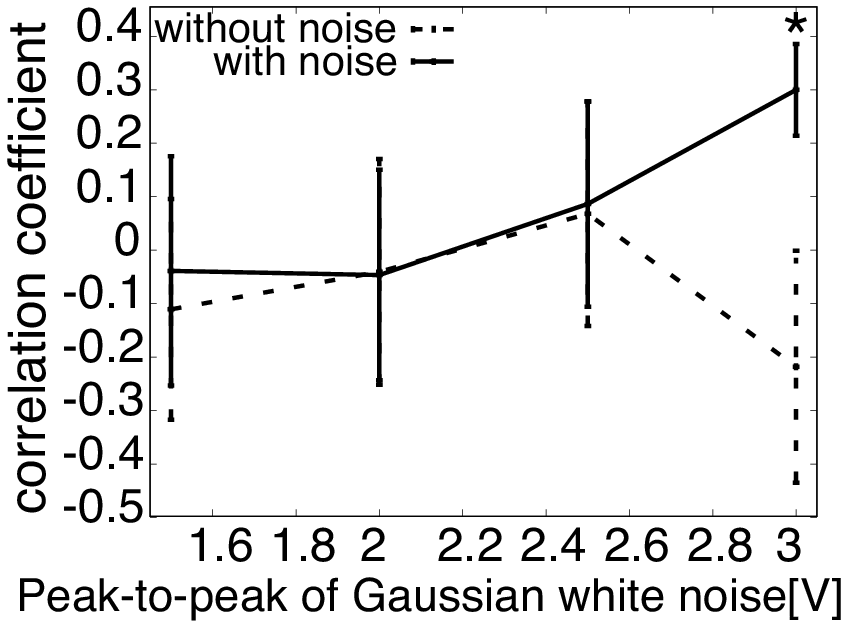}
(c)\includegraphics[scale=0.28]{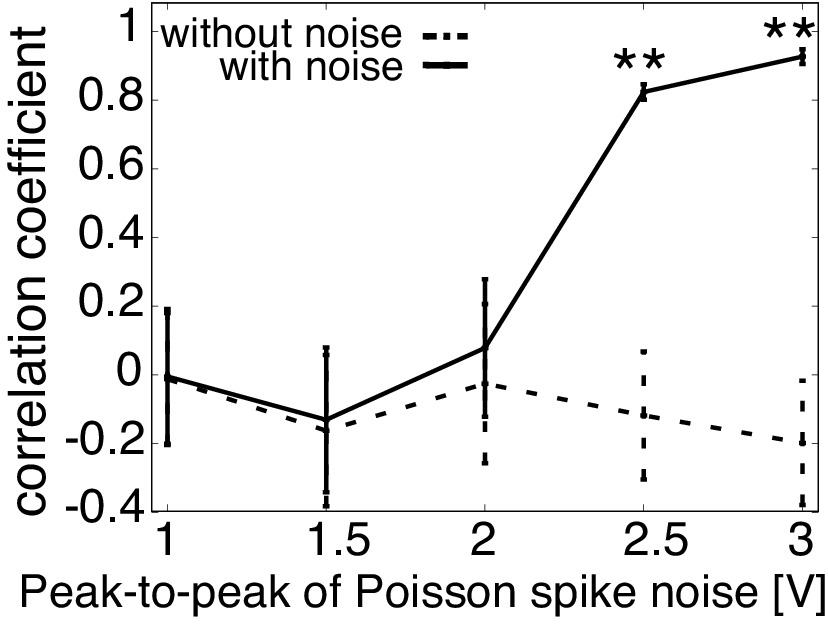}
\caption{
(a) Schematic diagram of a single \textit{Pierce} circuit with 
an external input supplied through the coupling capacitor. 
The function generator injects a same external signal 
(\textit{Gaussian} white noise or \textit{Poisson} spike trains) 
to the \textit{Pierce} circuit for 10 times repeatedly.
(b) Dependence of the correlation coefficient between the 
circuit outputs of different trials on peak-to-peak voltage 
of the white \textit{Gaussian} noise input. 
Averaged value over 45 pairs of different trials is plotted,
where the standard deviation is indicated by the error-bar.
The solid and dashed lines represent the cases during and before
the noise injection, respectively
($\ast$: $t$-test $p < 0.05$, $\ast\ast$: $p < 0.001$).
(c) Dependence of the correlation coefficient between the 
circuit outputs of different trials on amplitude of 
\textit{Poisson} spike inputs (average frequency: 20 kHz). 
}
\label{one_pierce_ext}
\end{figure}

To carry out the experiment described above, we implemented 
a single \textit{Pierce} oscillator circuit, which has the 
same parameter setting as in the previous Section 
(see Fig.~\ref{one_pierce_ext}~(a) and Table~\ref{piercec}).
The function generator was composed of an embedded computer 
(Raspberry Pi 2 Model B) and a D/A converter (Sunhayato MM-5102), 
which is capable of generating a same sequence of white 
\textit{Gaussian} noise or \textit{Poisson} spike trains as 
an audio signal with a sampling rate of 192 kHz.

For each trial, output from the circuit and the noise signal were
recorded simultaneously. 
To align the time sequences of different trials, the time shift
that maximized the correlation coefficient between the noise signals
was sought. By this alignment, all trials give rise to the same
timing of noise inputs and thus the circuit outputs can be regarded 
as those receiving the same noise input simultaneously. 
Then, to quantify the level of synchrony, correlation coefficient 
between the circuit outputs of different trials was computed. 

Fig.~\ref{one_pierce_ext} (b) shows the results of applying 
\textit{Gaussian} white noise.
Peak-to-peak voltage of the noise signals was increased from 1.5 
to 3 V. For each noise intensity, 10 trials with the same noise
signal were repeatedly made.
For 45 pairs of all possible combinations, the correlation coefficients 
between the circuit outputs were computed. Their average was drawn 
by the solid line, while the standard deviation was indicated by 
the error-bar. We see that the output coherence increases slightly 
as the noise intensity is increased.
To compare with the case that no noise was injected, the correlation 
coefficient, which was computed in a similar manner between the 
circuit outputs \textit{before} the noise was injected, was
plotted by the dotted line.
The significant difference (${\ast}$: $t$-test, $p < 0.05$) 
was detected for the noise intensity of 3 V, indicating that the 
observed increase in the level of output coherence was due to 
the common \textit{Gaussian} white noise.

Fig.~\ref{one_pierce_ext} (c), on the other hand, shows the results 
of applying \textit{Poisson} spike trains (mean frequency: 20 kHz).
Again, the correlation coefficient between the circuit outputs 
increased clearly as the noise intensity was increased.
For the noise intensity of 2.5 V and 3 V, significant difference 
(${\ast}{\ast}$: $t$-test, $p < 0.001$) from the control signals 
was detected, indicating that the oscillator outputs were highly 
correlated with each other due to the common spike inputs. 

\begin{figure}[t!]
\centering
(a)\includegraphics[scale=0.28]{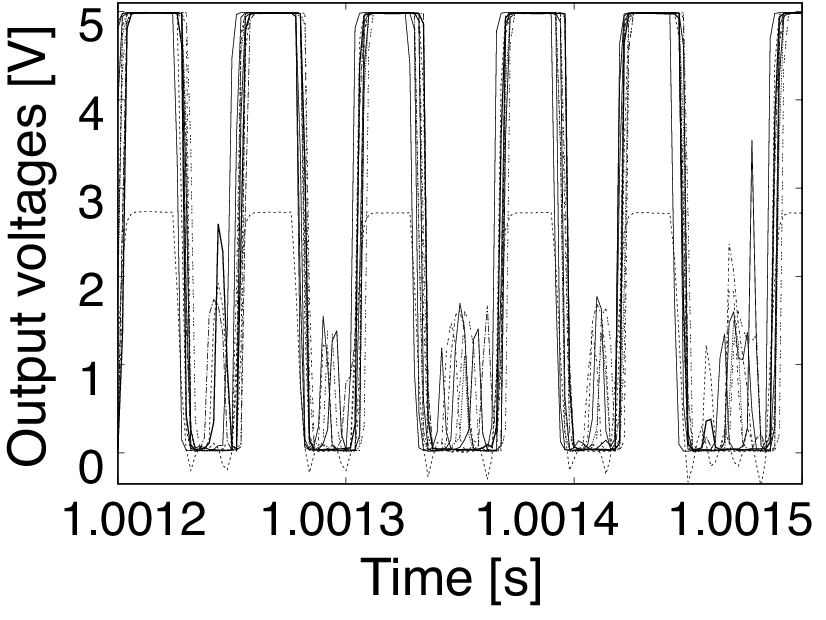}
(b)\includegraphics[scale=0.28]{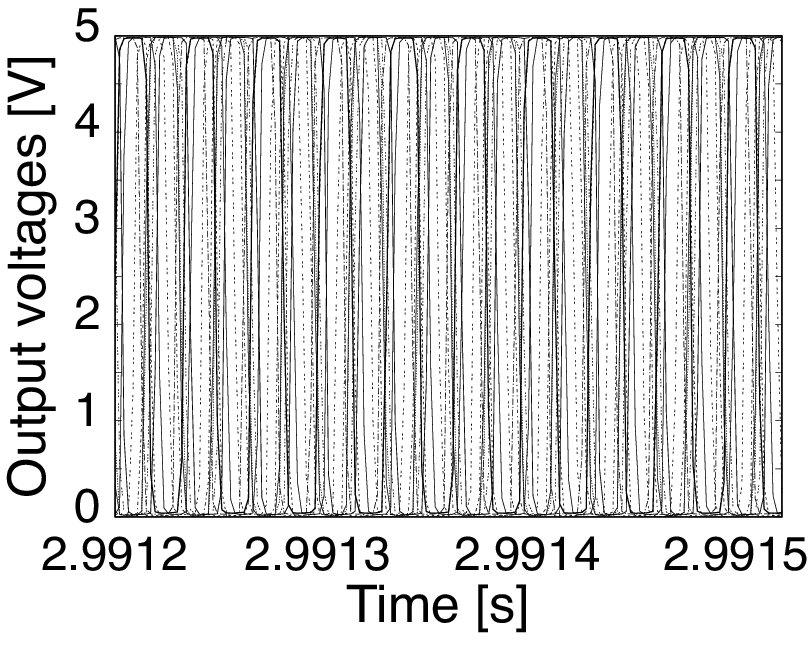}
(c)\includegraphics[scale=0.2]{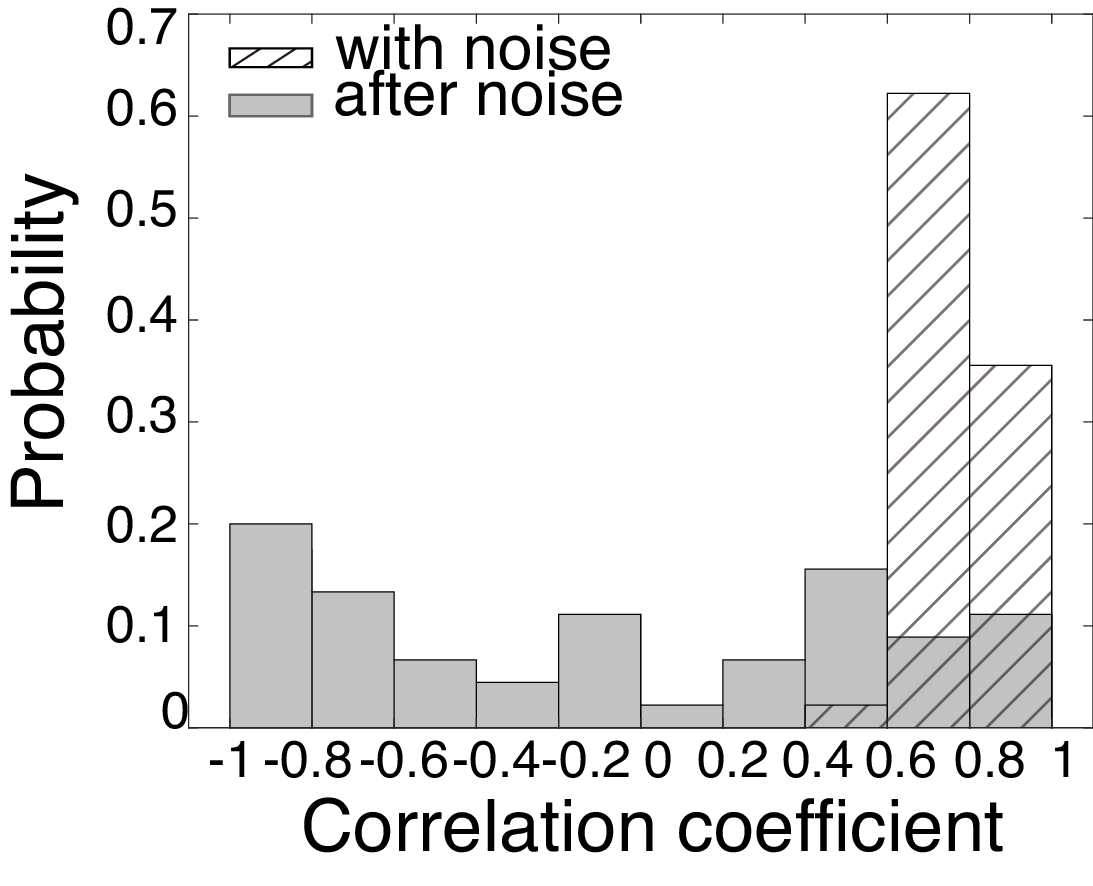}
\caption{
(a) Simultaneous plots of the output signals ($V_{1}$) 
recorded for 10 times with an injection of same \textit{Poisson} 
spike trains (mean frequency: 20 kHz, spike voltage: 3 V).
(b) As in (a), simultaneous plots of output signals drawn 
after the \textit{Poisson} spike input was terminated. 
(c) Histogram of the correlation coefficients between the 
circuit outputs computed for 45 pairs of different trials. 
The stripe and gray patterns correspond to the cases 
during and after the injection of \textit{Poisson} spike 
trains, respectively.
}
\label{one_ts_histo}
\end{figure}

To see the post effect of the noise-induced coherence, 
the circuit outputs were further examined after the noise 
injection was terminated. For \textit{Poisson} spike trains 
(mean frequency: 20 kHz, spike voltage: 3 V), the output signals 
($V_{1}$) were drawn simultaneously for 10 different trials.
Fig.~\ref{one_ts_histo} (a) and (b) correspond to the plots
during and after the noise injection, respectively. 
Strongly correlated outputs discernible during the noise 
injection quickly disappeared after the noise injection. 
To quantitatively see the change in the correlation, 
Fig.~\ref{one_ts_histo}(c) displays distribution of the 
45 correlation coefficients computed for all pairs of the 
10 trials during (stripe patterns) and after (gray patterns) 
the noise injection. 
The high correlation, which existed during the noise injection,
disappeared after the noise injection was stopped. 
According to the Kolmogorov-Smirnov test, their difference was
significant ($p=0.0069$).

These results suggest that, 
in our experiment of repeating the same noise injection to
the single \textit{Pierce} circuit, 
the noise input increased coherence of the oscillator outputs
by modulating the circuit system. 
This noise perturbation was, however, not strong enough to 
induce \textit{phase shift} of the crystal oscillators, since 
the coherence was lost instantly after the noise injection 
was stopped.
To induce phase synchronization of limit cycle oscillators,
the phase shift, which lasts permanently after the perturbation, 
is required
\cite{winfree2001,kuramoto1984,pikovsky2003synchronization}.
In this sense, the present experimental framework is not sufficient
to realize noise-induced synchronization of the crystal oscillators.

\section{Voltage Resetting as the Strongest Perturbation}
\begin{figure}[t!]
\centering
(a)\includegraphics[scale=0.45]{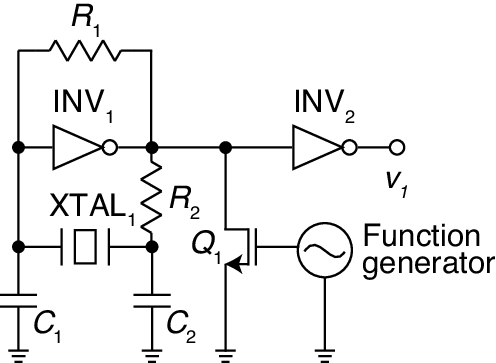}
(b)\includegraphics[scale=0.23]{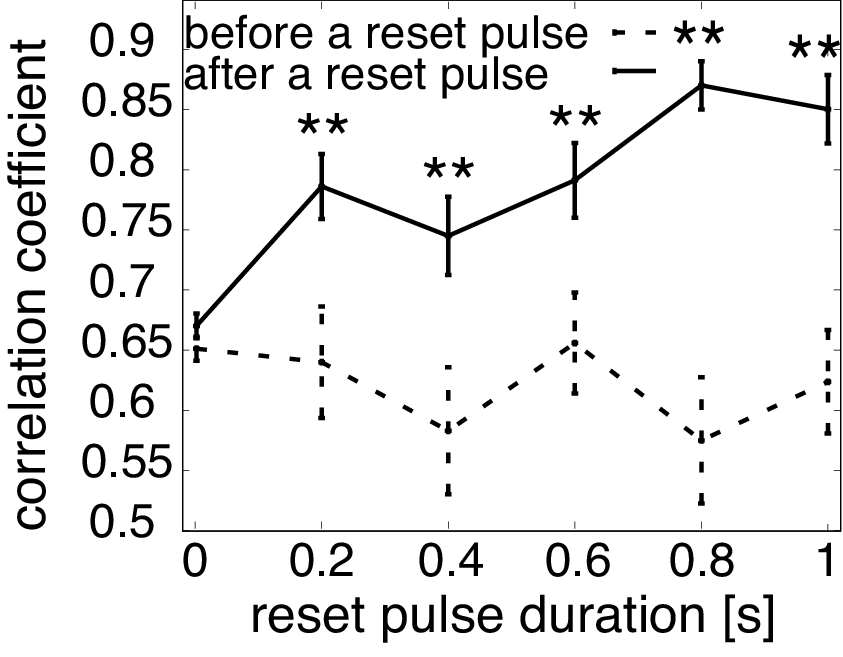}
(c)\includegraphics[scale=0.28]{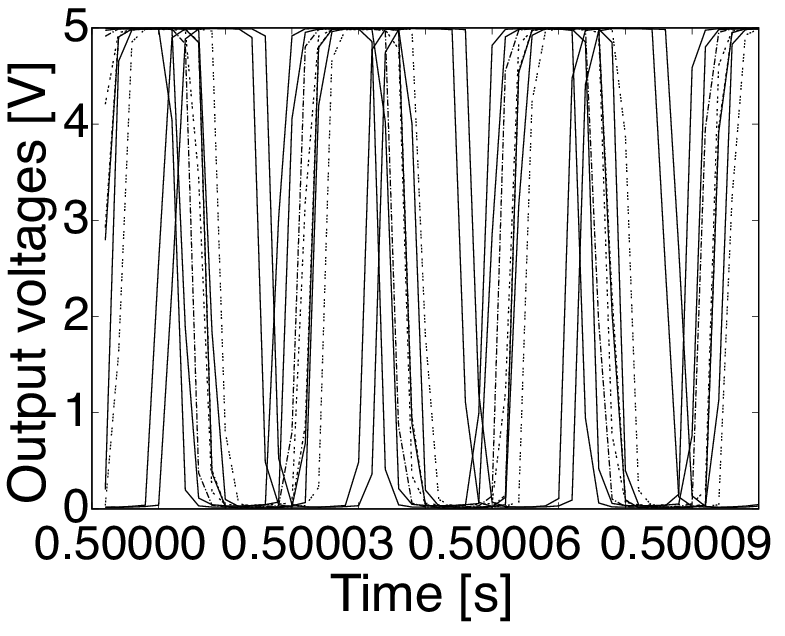}
(d)\includegraphics[scale=0.28]{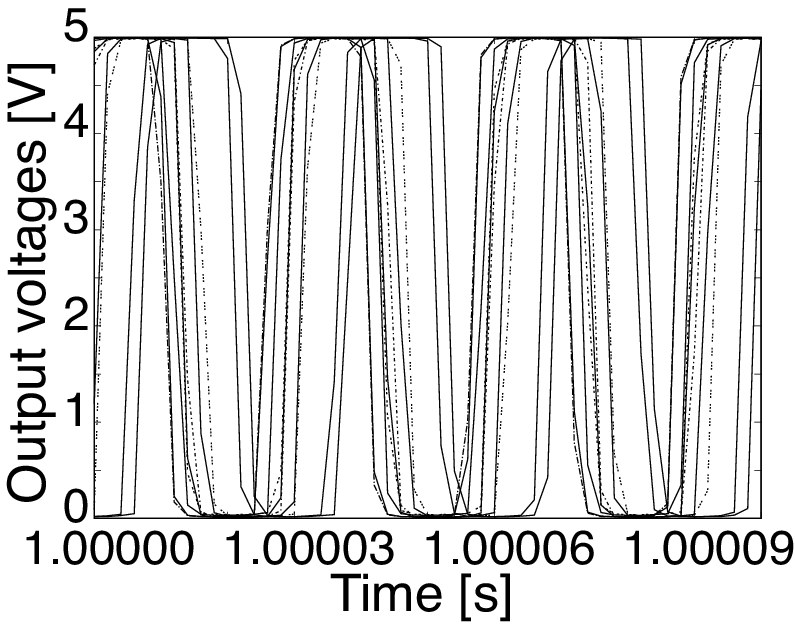}
(e)\includegraphics[scale=0.28]{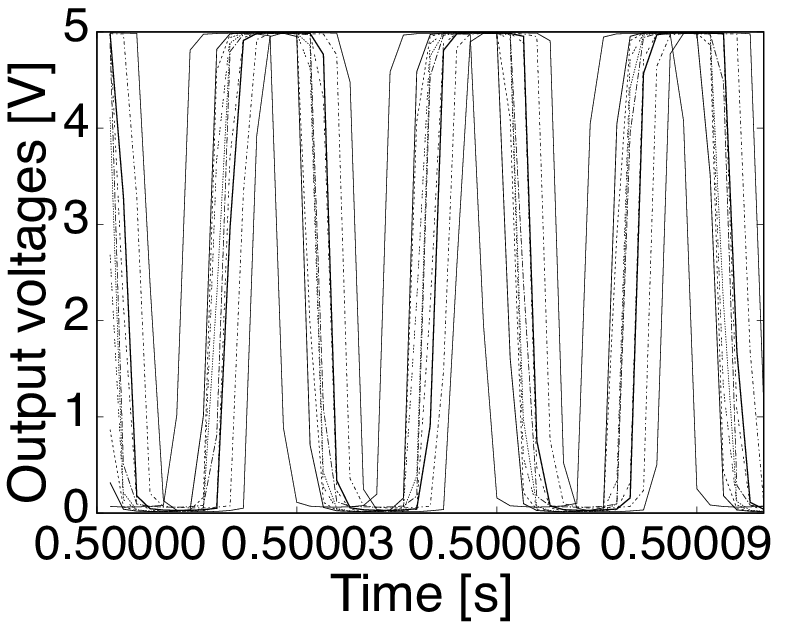}
(f)\includegraphics[scale=0.28]{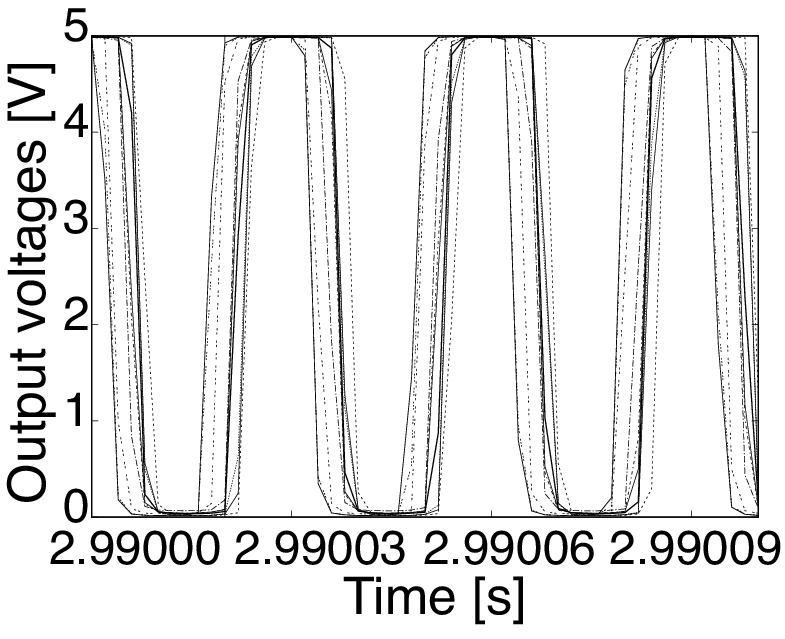}
\caption{
(a) Schematic diagram of a single \textit{Pierce} oscillator circuit. 
The function generator transmits a reset signal to the
nMOS transistor (${\rm Q_1}$).
When ${\rm Q_1}$ is turned on, the output voltage is reset 
from ${\rm INV_1}$ to 0 V. 
(b) Dependence of the correlation coefficient between output 
signals of different trials on the resetting time duration. 
The dashed and solid lines represent the cases before and after 
the injection of the resetting signal, respectively
(${\ast}{\ast}$: $p < 0.001$, $t$-test).
The error-bars represent standard deviation over all pairs of
10 trials (for reseting time of 2 ms, 50 trials were made). 
(c-f) Simultaneous plot of the output signals $V_{1}$ for 10 different 
trials before (c,e) and after (d,f) the voltage resetting. 
The resetting time was set to 2 ms (c,d) and 0.4 s (e,f). 
The waveforms were sorted so that the reset timings become the same.
}
\label{one_pierce_rst}
\end{figure}

Our experiment in the previous Section indicated that 
the repeated injection of the same noise to the single 
\textit{Pierce} circuit
was not strong enough to induce the real phase shift, which 
should last permanently after the noise input was terminated. 
As much stronger stimulus to induce such phase shift, we introduced 
a voltage resetting of the \textit{Pierce} oscillator circuit 
through the nMOS transistor (${\rm Q_1}$) as shown in 
Fig.~\ref{one_pierce_rst}~(a).
The function generator, used in our previous experiment as
Fig.~\ref{one_pierce_ext}~(a), transmitted a resetting signal of 
high voltage to the transistor ${\rm Q_1}$.
By this signal, the output voltage was reset from ${\rm INV_1}$ 
to 0 V, during the time when the transistor ${\rm Q_1}$ was 
turned on. 

In our first experiment, the resetting time, during which the 
output voltage was reset to 0 V, was set to 2 ms. 
In each trial, a single resetting signal was injected to the 
\textit{Pierce} circuit. Then, the resetting signal and the 
output signal were recorded simultaneously.
This measurements were repeated for 10 times. 
As in our previous experiment, to align the time sequences of 
different trials, the time shift that maximized the correlation 
coefficient between the resetting signals was sought. 
By this alignment, all trials give rise to the same timing 
of resetting signals and thus the circuit outputs can be regarded 
as those simultaneously reset.
Then, to quantify the level of synchrony, correlation coefficient 
between the circuit outputs of different trials was computed. 

Fig.~\ref{one_pierce_rst} (c) and (d) show simultaneous plots 
of the output signals ($V_{1}$) with 10 different trials.
No clear difference is observed between the plots before (c) 
and after (d) the input of resetting signal.
Fig.~\ref{one_pierce_rst} (e) and (f), on the other hand, show 
the case that the resetting time was set to 0.8 s. 
The phase shifts were induced clearly after the resetting in (f), 
showing a sign of synchronization among the outputs of different 
trials. These phase shifts were not instantaneous but they 
lasted for a very long time after the voltage reset.

To quantify the level of synchrony, correlation coefficient 
between the circuit outputs of different trials was computed
for all pairs of different trials.
Fig.~\ref{one_pierce_rst} (b) shows dependence of the averaged
correlation coefficient on the resetting time duration. 
The dashed and solid lines represent the cases before and after 
the reset input, respectively. 
For resetting time longer than 0.2 s, significant increases 
(${\ast}{\ast}$: $p < 0.001$, $t$-test)
in the correlation coefficient are recognized.
Longer the resetting time is, more coherent the output signals
become. This is reasonable because the resetting effect becomes 
stronger as the resetting time is extended, inducing a stronger
level of synchrony in the \textit{Pierce} oscillator circuit.

\section{Discussions and Conclusions}
The present study focused on the noise-induced synchronization 
of crystal oscillators.
Two uncoupled \textit{Pierce} circuits receiving a common noise 
input and a single \textit{Pierce} circuit repeatedly forced by 
a same noise input were implemented in our hardware.
Regardless of noise amplitude and noise types, synchronization 
was not achieved between the two uncoupled crystal oscillators.
Then, to exclude the frequency mismatch of the oscillators as 
a primary cause of avoiding the synchrony, a same noise input
was injected repeatedly to a single \textit{Pierce} circuit.
Although the coherence of the circuit outputs was increased
during the noise injection, the output coherence disappeared 
immediately after the noise injection, implying that the noise 
was not strong enough to induce the real phase shifts. 
As the strongest perturbation, the voltage resetting was finally 
examined. As the resetting time was increased, a clear phase 
shift was induced, giving rise to synchronized outputs of the 
circuit oscillator. 
Taken together, our study indicated that the crystal oscillator 
was robust against eternal perturbations such as noise injections 
in the sense that its phase was not easily shifted. 
Much stronger perturbation such as the voltage resetting was 
needed to induce the phase shifts, leading to synchrony. 

Recently, noise-induced synchronization has been applied to
several engineering problems. 
For instance, simulation study examined environmental noise 
as a possible source for synchronizing wireless sensor 
networks \cite{yasuda2013natural,yasuda2016synchronization}.
Noise-induced synchronization has been also utilized
in a simulated array of spin torque oscillators to 
overcome their low output power \cite{nakada2012noise}. 
Along the line of these works, synchronization of the crystal 
oscillators would provide a rich technological basis for 
unifying multiple CPUs, sensor networks, and other distributed 
clock devices. 
Towards such applications, our study provides a guideline 
for realizing noise-induced synchronization of the crystal 
oscillators, where design of a very strong signal is needed 
to achieve their synchrony. 
In our present experiment, a single resetting pulse has been 
applied to the crystal oscillator. The important future study 
would be to apply multiple resetting pulses
\cite{tass2007phase,fukuda2013controlling}
with a short resetting time to induce more efficient phase 
shifts, leading to noise-induced synchronization of crystal 
oscillators. 

\section*{Acknowledgements}
The authors would like to thank Dr. Yoji Kawamura for 
stimulating discussions.
This work was partially supported by the 
Grant-in-Aid for Scientific Research 
(No. 17K14685, No. 16K00343, No. 16K06154, No. 26286086) 
from Japan Society for the Promotion of Science (JSPS). 


%

\end{document}